\documentstyle[epsf,aps,twocolumn,prl,floats]{revtex}  
\begin{document}
\draft
\title{Spontaneous vortex phase in ErNi$_{2}$$^{11}$B$_{2}$C}
\author {H. Kawano-Furukawa$^{1,2}$, E. Habuta$^{3}$, T. Nagata$^{1}$, M. Nagao$^{4}$, H. Yoshizawa$^{4}$, \\
 N. Furukawa$^{5}$, H. Takeya$^{6}$ and K. Kadowaki$^{7}$}
\address{$^{1}$Dept. of Physics, Ochanomizu univ., Bunkyo-ku, Tokyo 112-8610, Japan.}
\address{$^{2}$PRESTO, Japan Science and Technology Corporation, Kawaguchi, Saitama 332-0012, Japan.}
\address{$^{3}$G. S. H. S., Ochanomizu univ., Bunkyo-ku, Tokyo 112-8610, Japan.}
\address{$^{4}$Neutron Scattering Lab., I. S. S. P., Univ. of Tokyo, Tokai, Ibaraki 319-1106, Japan.}
\address{$^{5}$Dept. of Physics, Aoyama Gakuin Univ., Setagaya, Tokyo 157-8572, Japan.}
\address{$^{6}$National Institute for Materials Science, Tsukuba, Ibaraki 305-0047, Japan.}
\address{$^{7}$Institute of Materials Science, Univ. of Tsukuba, Tsukuba, Ibaraki 305-0006, Japan.}

\date {\today}

\twocolumn[\hsize\textwidth\columnwidth\hsize\csname @twocolumnfalse\endcsname

\maketitle
\begin{abstract} 
Below $T_{\rm WFM}$ $\sim$ 2.75K, a weak ferromagnetism and superconductivity coexist in ErNi$_{2}$B$_{2}$C. In the present paper, a possibility of a spontaneous vortex phase was examined directly by small angle neutron scattering technique. An external magnetic field was applied to align ferromagnetic domains so that the internal magnetic field induced by the weak ferromagnetism becomes uniform and uniaxial over the bulk sample. When the field is removed, the flux line lattice subsisted below $T_{\rm WFM}$ but disappeared above it. A pinning effect is also discussed. We conclude that the spontaneous vortex state is realized in ErNi$_{2}$B$_{2}$C.

\end{abstract}
]


Ferromagnetism (FM) and superconductivity (SC) are both cooperative quantum phenomena which occur at $q=0$, but a widely accepted consensus is that they are mutually exclusive.  The pioneering theory on this issue by Ginzburg \cite{Gin57} stimulated a number of theoretical works which follow continuously since 1950's up to now.  At present, the coexistence of such two states is believed to occur only when an internal magnetic field $H_{\rm int}$ induced by a spontaneous FM moment is smaller than the critical magnetic field $H_{\rm c}$ for a type I SC and  $H_{\rm c2}$ for a type II SC, respectively\cite{Fis90}.  Moreover, for a type II SC, an appearance of a spontaneous vortex phase (SVP) is predicted if $H_{\rm int}$ further satisfies a condition, $H_{\rm c1}< H_{\rm int} < H_{\rm c2}$\cite{Svp1,Svp2,Svp3}. 
By contrast, intensive experimental studies on FM SCs were activated in the late 1970's by the discovery of two families of magnetic intermetallic SCs, RE-Rh$_{4}$B$_{4}$ and RE-Mo$_{6}$X$_{8}$ (RE=rare earth)\cite{Fis90}.  Some compounds in these two families were, indeed, confirmed to show both FM and SC.  When the FM ordering was established in these systems, however, SC was immediately destroyed, and the coexistence of these two states was observed only in a very limited narrow range of temperature like $\Delta T \sim 0.1\rm K$ just below $T_{\rm C}$\cite{Mon80,Lyn81}.

Recently, the study of magnetic SCs has been revived due to the discovery of two {\it new} magnetic SCs. One is an intermetallic SC,  RETM$_2$B$_2$C (TM=Ni, Pd) \cite{Boro1,Boro2,Boro3}, while the other is a family of Rutheno-cuprate compounds, RuSr$_2$RE$_2$Cu$_2$O$_{10}$ \cite{R1222} and RuSr$_2$RECu$_2$O$_8$ \cite{R1212a,R1212b,R1212c}. By magnetization, specific heat and/or resistivity measurements, the coexistence of weak ferromagnetism (WFM) and SC was suggested for ErNi$_2$B$_2$C \cite{Can96} as well as Rutheno-cuprates \cite{R1222,R1212a,R1212b,R1212c}, and a possibility of the SVP on these compounds was theoretically discussed\cite{Ng97}. A couple of experimental reports concluded the existence of the SVP in Rutheno-cuprates \cite{SVP4,SVP5}, but such conclusion should be taken with caution because it was based on only macroscopic measurements. In fact, very recent neutron scattering studies proved that RuSr$_2$RECu$_2$O$_8$ (RE = Gd, Y) has actually an antiferromagnetic ground state, and under an applied field, the {\it field-induced} WFM order coexists with SC \cite{Lyn00,Tak00}.   From this lesson, we would like to emphasize that it is crucial to confirm coexistence of FM and SC by means of microscopic measurements. 


ErNi$_2$B$_2$C is the first material which is microscopically confirmed to exhibit the coexistence of WFM and SC below $T_{\rm WFM}$ $\sim$  2.75 K\cite{Kaw99,Fur012}.  The results of detailed analyses of the magnetic structure in the WFM phase will be reported elsewhere\cite{Fur01}. Note that the previous neutron scattering studies revealed that localized magnetic moments on Er atoms first form a transversely-polarized planar-sinusoidal (SDW) structure below $T_{\rm N}$ $\sim 6$ K with an incommensurate propagating vector $q_{1} \sim 0.553 a^{*}$ and an amplitude of the modulated moment $\mu = 7.8 \mu_{\rm B}$ \cite{Zar95,Sin95}. The crystal structure of ErNi$_2$B$_2$C belongs to tetragonal I4/mmm symmetry at room temperature, but it transforms to an orthorhombic phase below $T_{\rm N}$\cite{Det97}.  The magnetic moments in the ordered state are parallel to the orthorhombic $b$-axis\cite{Det00}.

To further examine the influence of the WFM ordering to SC, we have studied temperature ($T$) dependence of lower critical field $H_{\rm c1}$ in details\cite{Fur012}.  As explained in the inset of Fig. 1, three different magnetization curves, ( (1) vir : virgin curves, (2) dec : H-decreasing curves, and (3) inc : H-increasing curves) were measured with applying an external magnetic field $H_{\rm ext}$ either parallel to the $b$ ($\parallel b$) or $c$  ($\parallel c$) axes, and the corresponding  $H_{\rm cr}$ were determined for each curves as the field where the two guided lines are crossing as exemplified in the inset.  The difference between  $H_{\rm c1}$ and $H_{\rm cr}$ is very small. Therefore, one can approximetly regard  $H_{\rm cr}$ as $H_{\rm c1}$. (For detailed explanations, see Ref. \onlinecite{Fur012}.) It should be noted here that a magnetization curve of the magnetic SC is inevitably affected by $H_{\rm int}$ produced by the magnetic components.  For $H_{\rm ext}  < H_{\rm c1}$, however, the influence of $H_{\rm int}$ is expected to be very weak because the  spins which contribute $H_{\rm int}$ are limited to those which reside within the penetration depth $\lambda$ from the surface, thereby justifies the following analyses and discussions.

\begin{figure}
\centering \leavevmode
\epsfxsize=0.75\hsize
\epsfbox{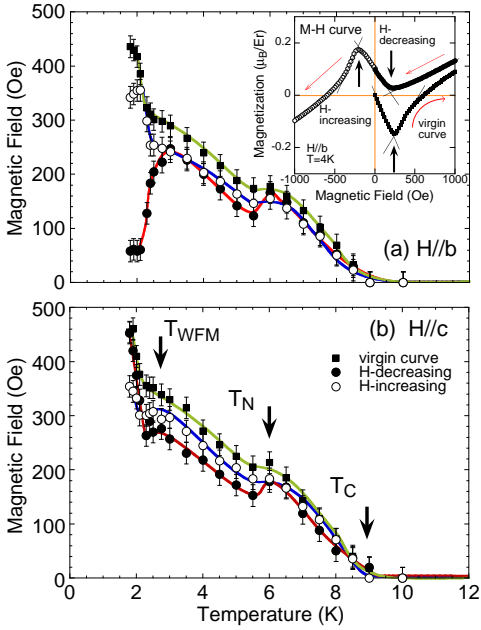}
\vspace{2mm}
\caption{(color)Temperature dependence of lower critical field: (a)$H_{\rm ext}^{\parallel b}$, and (b)$H_{\rm ext}^{\parallel c}$, respectively. The inset of the top panel shows a typical example of $M-H$ curves observed at $T = 4$ K for $H_{\rm ext}^{\parallel b}$. After Ref. [22].}
\label{Fig.1}
\end{figure}

In Fig. 1 is depicted thus obtained $T$ dependence of $H_{\rm cr}$.  The upper and lower panels correspond to behavior of $H_{\rm cr}$ for $H_{\rm ext}^{\parallel b}$ and $H_{\rm ext}^{\parallel c}$, respectively\cite{Fur012}. All six curves exhibit a weak dip near $T_{\rm N}$. They are attributed to the magnetic SDW and a structural phase transition at $T_{\rm N}$.  By contrast, the drastic influence was observed at $T_{\rm WFM}$.  $H_{\rm cr}$ determined by $M_{\rm vir}$ curves with $H_{\rm ext}^{\parallel b}$ shows a precipitous decrease, whereas all other curves show an abrupt increase.  By a field-cycling procedure, we confirmed that these behaviors are completely reproducible.  Accordingly, the drastic changes of $H_{\rm cr}$ below $T_{\rm WFM}$ must be intrinsic to ErNi$_{2}$B$_{2}$C.

The mechanism of the decrease of $H_{\rm cr}$ below $T_{\rm WFM}$ can be interpreted as follows.  Remember that the decrease of $H_{\rm cr}$ was observed only in the $M_{\rm dec}$ curve for $H_{\rm ext}^{\parallel b}$, but not observed for $H_{\rm ext}^{\parallel c}$.  This distinct difference can be understood by taking accounts of the domain effect and of the magnetic anisotropy of ErNi$_{2}$B$_{2}$C simultaneously. ErNi$_{2}$B$_{2}$C has a strong basal plane magnetic anisotropy and, as aforementioned, it undergoes the tetragonal-to-orthorhombic structural transition at $T_{\rm N}$. Hence, after the zero field cooled (ZFC) procedure, there are four-fold magnetic and crystallographic domains below $T_{\rm N}$ within the basal plane.  Below $T_{\rm WFM}$, the WFM order naturally forms a domain structure, and the internal field $H_{\rm int}$ produced by the spontaneous FM component would be canceled out to zero.  It explains why the system shows a complete Meissner effect in all $M_{\rm vir}$ curves. In another words, after the ZFC procedure and for $H_{\rm ext} < H_{\rm c1}$, the system behaves just like it does not know the existence of the WFM ordering.  When $H_{\rm ext}$ is further increased and exceed the $H_{\rm c1}$, it starts to penetrate into the sample. Here, if the $H_{\rm ext}$ is parallel to the $b$-axis, it starts to align not only magnetic domains but also crystallographic domains. Once FM domains are aligned, $H_{\rm int}$ remains finite and subsists with keeping parallel to $H_{\rm ext}$ until $H_{\rm ext}$ decreases to 0 T.  The finite $H_{\rm int}$ supplements $H_{\rm ext}$, and $H_{\rm c1}$ ($H_{\rm cr}$ in the present study)  is drastically decreased below $T_{\rm WFM}$.  Because of a strong in-plane anisotropy, however, $H_{\rm ext}^{\parallel c}$ cannot align magnetic domains, and a decrease of $H_{\rm cr}$ is not observed.

We interpreted that the increase of H$_{\rm cr}$ below $T_{\rm WFM}$ indicates a possibility that the WFM enhances the SC Meissner phase.  Since the WFM coexists with SC, it may not be intuitively unreasonable to suppose that the formation of the WFM microscopically enhances a stability of the SC phase.  If the WFM favors the SC, a new interaction must be invoked between localized FM moments and conduction electrons\cite{PJ00}.

 The most important result here is that, at lower $T$'s, $H_{\rm cr}$ determined by the $M_{\rm dec}$ curves with $H_{\rm ext}^{\parallel b}$ seems to go to  practically zero with decreasing $T$.  Since the vortex phase is expected to exist in the field range for $H_{\rm ext} > H_{\rm c1}$, the present result strongly suggests a possibility that, when the $H_{\rm ext}$ is reduced from high field, the vortex state remains even $H_{\rm ext}$ is reduced to 0 T, which should be regarded as SVP by definition. In what follows, we will report the direct observation of the flux line lattice (FLL) by small angle neutron scattering (SANS) technique.



We used the same Boron-11 substituted single crystals of ErNi$_2$$^{11}$B$_2$C  which were used in our previous studies\cite{Kaw99,Fur012}.   Note that the $T_{\rm c}$, $T_{\rm N}$ and $T_{\rm WFM} $ of our sample were determined to be 8.6 K, 6.0 K and $\sim 2.75$ K, respectively\cite{Com01,Note1}. The remnant magnetization at 1.8 K was $\sim$ 0.34$\mu_{B}$/Er.  SANS experiments were performed by using  the SANS-U spectrometer at the C1-2 port in the JRR-3M guide hall, JAERI, Tokai, Japan. The B and Er have high neutron absorption factor. In order to reduce the absorption of neutrons,  we substitute natural B with $^{11}$B isotope(99 at.$\%$), but we could not use Er isotope. Therefore, to enhance the transmission of cold neutrons, we needed to slice the large single crystals into pieces with dimensions of $4 \times 1 \times 10$mm$^{3}$. Incident neutrons with wavelength $\lambda _{\rm n}$ = 11 \AA ~were selected by a velocity selector and detector position after sample was 16 m.  The sample was mounted in the cryostat with a horizontal magnet with a field direction and incident beam to be parallel to the sample $b$ axis. The data are collected by a 128 $\times$ 128 pixelated 2D position sensitive detector.  Because of low intensity,  it took 12 hours to obtain one FLL pattern. 
First the sample was cooled in zero field (ZFC).  In order to align the magnetic domains,  $H_{\rm ext}$ of 2.8 T  is applied parallel to the $b$ axis and it is removed. The FLL measurements are repeated at 1.6 K below $T_{\rm WFM} \sim 2.75$ K and at 4.0 K above $T_{\rm WFM}$, respectively.  Our idea is, if SVP is realized in the WFM phase, then we should be able to observe the FLL signal at 1.6K but not at 4.0 K when $H_{\rm ext}$ is reduced to 0 T.


\begin{figure}
\centering  \leavevmode
\epsfxsize=0.79\hsize
\epsfbox{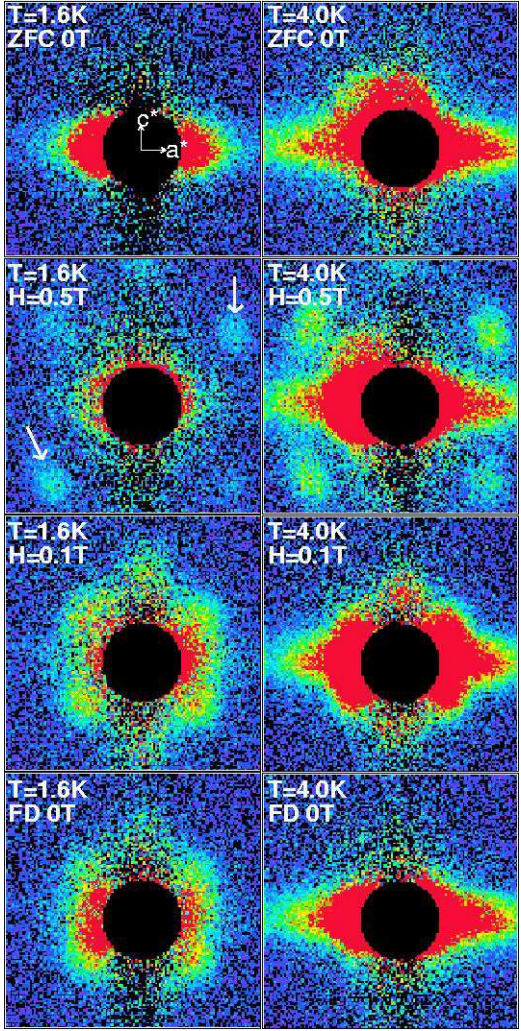}
\vspace{1mm}
\caption{(color) FLL diffraction  patterns at 1.6 K (left column) and 4.0 K (right column) observed for ZFC0T, 0.5T, 0.1T and FD0T procedures from the top to bottom panels, respectively. Here, ZFC and FD denote "zero-field-cooled" and "field-decreased", respectively.  A background data taken at $T = 10$ K was subtracted. The crystalline axes are shown in the left-top panel. }
\label{Fig.2}
\end{figure}

Figure 2 shows typical FLL diffraction patterns in ErNi$_{2}^{11}$B$_{2}$C. As a background the 10 K data in the paramagnetic normal state was subtracted\cite{Com02}. To omit the strong background scattering, the data in a small angle region (in radius range of r $<$ 20 pixel) are masked in the figures. The sample crystallographic orientation is indicated in the left-top panel. At 0 T after the ZFC procedure (ZFC0T), shown in the top 2 panels, only FM diffuse scattering is observed along the $a^{*}$ axis in a small $q$ region. After applying $H_{\rm ext}$ as large as 2.8 T, it was reduced down to 0.5 T, and then  to 0.1 T (middle 4 panels). The diffuse scattering observed at ZFC0T is strongly suppressed at 1.6 K but not at 4.0 K, demonstrating that magnetic domains are aligned by $H_{\rm ext}$ at 1.6 K. In addition, the FLL signal is clearly observed in all four cases.  We noticed that the $q$ position for the FLL signal at 1.6 K is larger than that at 4.0 K, indicating that the system feels an extra magnetic field at 1.6 K. This is consistent with our idea that $H_{\rm int}$ by WFM supplements $H_{\rm ext}$ and affects the position of vortices. 
The FLL patterns after field decreasing to 0 T (FD0T) are depicted in the bottom 2 panels. At 1.6 K the FLL signal is still visible but it is missing at 4.0 K.

\begin{figure}
\centering  \leavevmode
\epsfxsize=0.75\hsize
\epsfbox{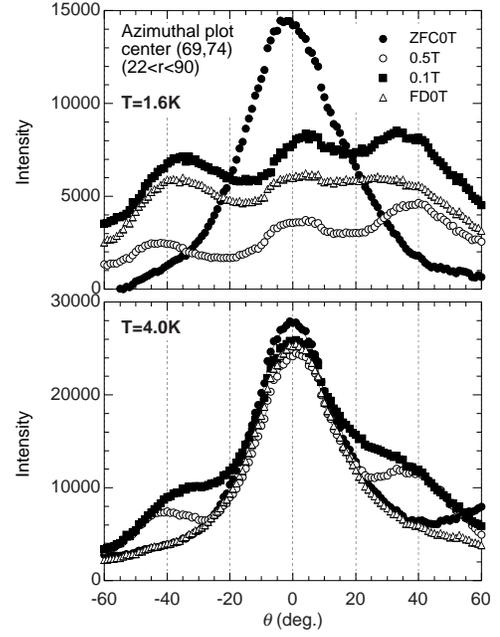}
\vspace{2mm}
\caption{Azimuthal dependence of FLL intensity. Data between $22<r<90$ pixel are used for the integration. (a) $T = 1.6$ K and (b) $T = 4.0$ K, respectively. }
\label{Fig.3}
\end{figure}


To further demonstrate the FLL signals, a portion of azimuthal dependence of scattering intensity is depicted in Fig. 3. Data in a radius range between $22<r<90$ pixel are used for this integration and $\theta$ is an angle from the $a^{*}$ axis.  For ZFC0T, only the diffuse components along the $a^{*}$ axis were observed at around $\theta \sim  0$ deg. At 0.5 T and 0.1 T, however, the diffuse component was drastically reduced at 1.6 K, while subsists at 4.0 K. In addition to the diffuse components, clear peaks are observed at around $\theta \sim  \pm 40$ deg, indicating the existence of FLL signals. Furthermore, for FD0T, the  FLL peaks remain around $\theta \sim  \pm 40$ deg at 1.6 K but completely disappear at 4.0 K, indicating that the vortices are swept out from the sample at 4.0 K at least until the FLL signals disappear from the experimental $q$ range.

Here we need to discuss the pining effect.  Recently, Dewhurst {\it et al.} pointed out that the bulk pining effect becomes important below $T_{\rm N}$ by using a miniature local Hall array\cite{Dew00}. Furthermore, Gammel {\it et al.} have demonstrated that a pinning force increases rapidly in the WFM phase by transport and magnetization measurements\cite{Gam00}. These results indicate a possibility that the enhanced pining mechanism below $T_{\rm WFM}$ may trap vortices, but the comparatively weak pining force above $T_{\rm WFM}$ would release them, hence the FLL signals were observed only at 1.6 K (below $T_{\rm WFM}$) but not at 4.0 K (above $T_{\rm WFM}$).  They argued, however, that the enhancement of the pining force was observed not only in the in-plane direction but also along the $c$ axis, and they were of comparable order.  This behavior has an important implication to interpret our results. Namely, in our study,  the decrease of $H_{\rm cr}$ was observed only with $H_{\rm ext}^{\parallel b}$ but not with $H_{\rm ext}^{\parallel c}$, exhibiting a clear anisotropy.  This result strongly indicates that the decrease of $H_{\rm cr}$ is due to the existence of $H_{\rm int}$ but not due to the pining effect. To give further confidence, one may expect that a similar experiment with $H_{\rm ext}^{\parallel c}$ would be valuable. The previous FLL measurements with this field configuration, however,  showed that the FLL rotate away from the applied magnetic field direction below $T_{\rm WFM}$\cite{Yar96}, which implys that  it is impossible to observe the FLL with a small $H_{\rm ext}^{\parallel c}$.  Based on these considerations, we conclude that the pinning effect is not essential to the present case, and that the SVP is realized in ErNi$_{2}$B$_{2}$C.

\begin{figure}
\centering  \leavevmode
\epsfxsize=0.70\hsize
\epsfbox{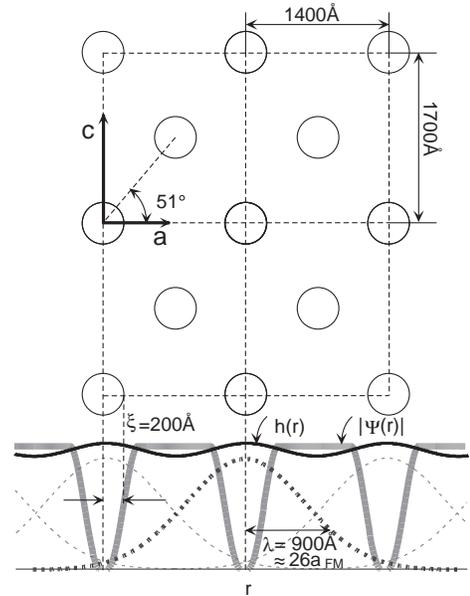}
\vspace{2mm}
\caption{Schematical picture of the vortex arrangement in the real space at 1.6 K in the SVP state for the FD0T procedure. }
\label{Fig.4}
\end{figure}

Finally in Fig.4 is depicted schematical arrangements of vortices in the real space at 1.6 K for the FD0T procedure. The field which corresponds to the observed $q$ position is calculated to be about 1600 G. The FLL is oriented with respect to the nearest neighbor direction parallel to the crystalline $a$ axis, being consistent with the previous work at 4 K with a lower field by Eskildsen {\it et al.}\cite{Esk01}.  From the values of $H_{\rm c1}$ ($H_{\rm cr}$ in the present study) and $H_{\rm c2}$ for our sample, the coherence length $\xi$ and the penetration depth $\lambda$ at 2 K are roughly estimated to be 200 \AA  ~and 900 \AA, respectively. Here to calculate $\lambda$, we employed $H_{\rm cr}$ for $H_{\rm ext}^{\parallel b}$ and the $M_{\rm vir}$ curve, because for the case of $H_{\rm ext}^{\parallel b}$ and $M_{\rm dec}$, the observed $H_{\rm cr}$ is already reduced by $H_{\rm int}$ from the field the system feels. Reflecting lower $T_{\rm c}$ of our sample, the calculated values are slightly larger than those reported in references\cite{yyy,uuu,ttt}.  Furthermore, our detailed analyses on the magnetic structure in the WFM phase revealed that the FM moment appears on the $(100)$ plane with the periodicity of about 35 \AA\cite{Fur01}. The calculated $\lambda$ is about 26 times larger than the periodicity of the FM sheets. The relation between these length scales is also depicted in the figure.

In summary, the vortex state in the WFM phase of ErNi$_{2}$B$_{2}$C was directly examined by the SANS technique. With a single FM domain, we  observed FLL even at $H_{\rm ext}$=0, and conclude that the SVP is indeed realized in ErNi$_{2}$B$_{2}$C.

We are grateful to Dr. M. Yathiraj for fruitful discussions. H. F. was supported by a Grant-In-Aid for Encouragement of Young Scientists from the Ministry of Education, Culture, Sports, Science and Technology, Japan.


\begin{references}
\bibitem{Com01} T$_{c}$ of our sample is 8.6 K and it is about 2 K lower than the one obtained by the flux method. Very recently it was found that annealing process raises $T_{\rm c}$ up to 10.5 K. We interpreted that metallurgical microstructures in the crystal decrease $T_{\rm c}$ . However, the results in the present paper were taken with the un-annealed sample.
\bibitem{Com02}As explained in the text, distributions of the magnetic domains are drastically changed by applying a magnetic field.  Therefore we cannot use the zero field cooled 0T data (top 2 panels) as a background data.
\end{references}
\end{document}